\documentstyle[amssymb,aps,multicol,epsf]{revtex}

\begin{document}
\draft

\title{
How to generate a random growing network
}

\author{
S.N. Dorogovtsev$^{1, 2, \ast
}$,  
J.F.F. Mendes$^{1, \dagger}$, 
and 
A.N. Samukhin$^{2, \ddagger}$
}

\address{
$^{1}$ Departamento de F\'\i sica and Centro de F\'\i sica do Porto, Faculdade 
de Ci\^encias, 
Universidade do Porto\\
Rua do Campo Alegre 687, 4169-007 Porto, Portugal\\
$^{2}$ A.F. Ioffe Physico-Technical Institute, 194021 St. Petersburg, Russia 
}

\maketitle
   
\begin{abstract} 
We propose a construction procedure which generates a wide class of random evolving networks with fat-tailed degree distributions and an arbitrary clustering. This procedure applies the stochastic transformations of edges, which can be used as a basis of a real space renormalization group for evolving networks.  
\end{abstract}


\begin{multicols}{2}
\narrowtext


The basis of the network science (see recent reviews \cite{s01,ab01a,dm01c}) 
are construction procedures. The main of them are: 

\begin{itemize}

\item[(i)] the classical random graphs of Erd\"{o}s and R\'enyi \cite{er59}, 

\item[(ii)] the random graphs with a given degree distribution \cite{mr95}, 

\item[(iii)] small-world networks introduced by Watts and Strogatz \cite{ws98},  

\item[(iv)] networks, growing under the mechanism of preferential linking (including, as a particular case, random linking), which were introduced by Barab\'asi and Albert \cite{ba99}. 

\end{itemize}

From a physical point of view, networks can be equilibrium and non-equilibrium. Most real networks are non-equilibrium. These nets are characterized by strong correlations, including, often, strong clustering. 

\begin{figure}
\epsfxsize=55mm
\centerline{\epsffile{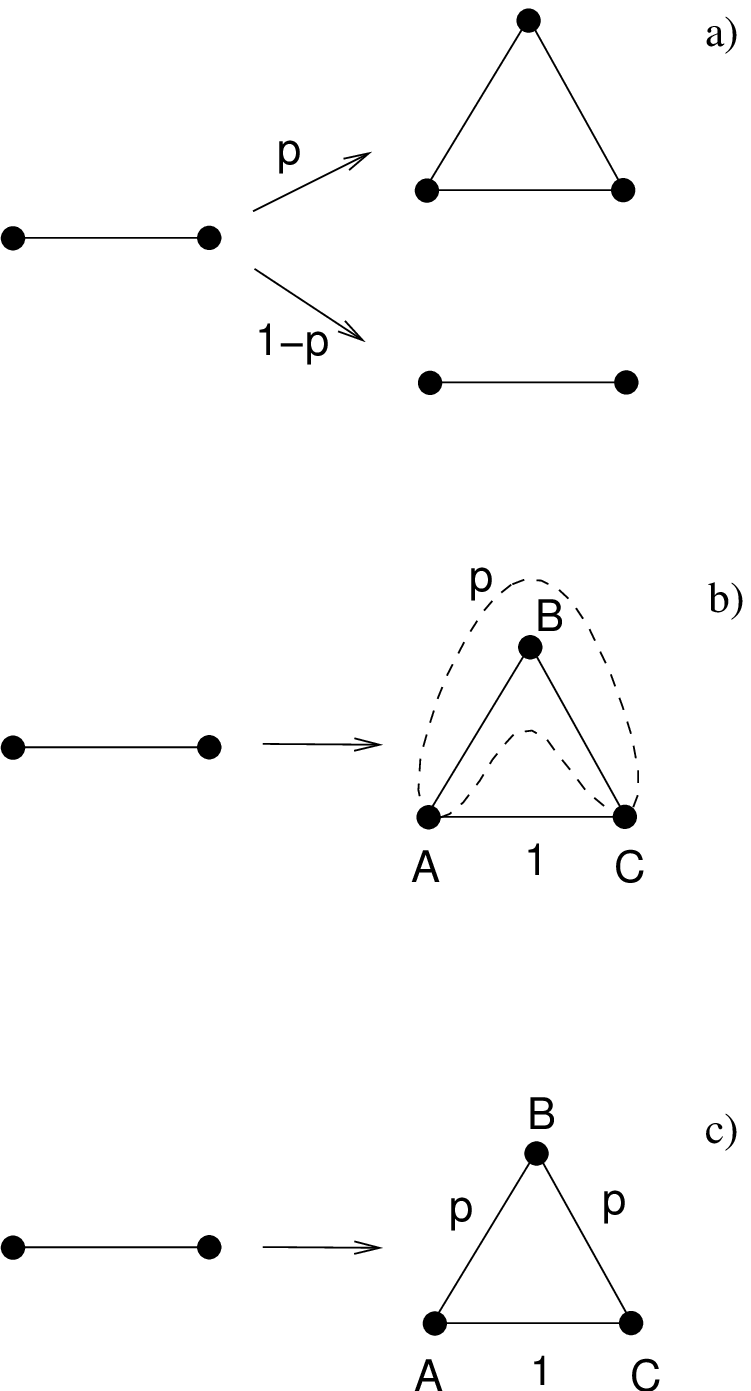}}
\caption{
Examples of the transformations of edges, which generate random growing networks.  
\newline 
(a) At each time step each edge of the network is transformed into one configuration of the two shown. $p$ and $1-p$ are the probabilities of these transformations. 
\newline
(b) Another representation of the same rule of the evolution of edges: the chain $ABC$ is present with the probability $p$. Particularly, if $p=1$, the transformation (a) or, equivalently, (b) generates the deterministic scale-free (strongly clustered) graph introduced in Ref. \protect\cite{dgm01}. For $p \to 0$ but $p \neq 0$, the transformation (a) or (b) generates the stochastically growing scale-free (strongly clustered) network introduced in Ref. \protect\cite{dm01c}. 
\newline
(c) A different transformation: at each time step each edge is transformed into the cluster, where edges $AB$, $BC$, and $AC$ are present with probabilities $p$, $p$ and $1$ respectively. If $p=1$, we again have the deterministically growing graph of Ref. \protect\cite{dgm01}. For $p \to 0$ but $p \neq 0$, the transformation (c) generates the Barab\'asi-Albert model.  
}
\label{f1}
\end{figure}

\begin{figure}
\epsfxsize=75mm
\centerline{\epsffile{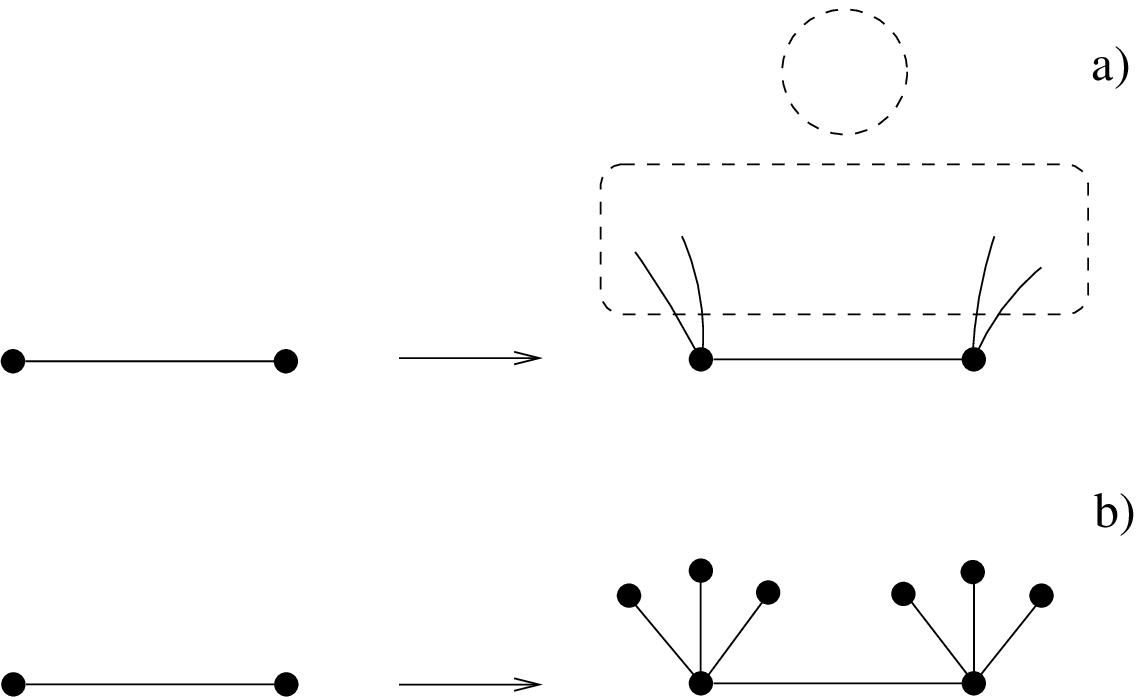}}
\caption{
The transformations, which generate growing graphs with a the small-world effect, may be of a more general form than in Fig. \protect\ref{f1}, see, for example, (a) where disconnected components emerge. In particular, transformation (b) generates scale-free trees (see Ref. \protect\cite{jkk02}).
}
\label{f2}
\end{figure}

Here we present a simple construction of random growing networks, which provides networks with fat-tailed degree distributions and an arbitrary clustering. 
In this construction, the evolution of a network is reduced to a stochastic recursive transformation of edges (more rigorously, a transformation of edges with their end vertices), that is, to the {\em evolution of edges}. A simple example of this transformation is shown in Fig. \ref{f1}(a). At each time step each edge of the network transforms into various configurations of edges and vertices with some probabilities. These configurations may be more complex and general than those which are shown in Fig. \ref{f1} (see Fig. \ref{f2}). 
Notice, that, generally, the resulting network may have disconnected components.  

Note that we only show the transformations that generate networks with the small-world effect. In the limiting case, in which an edge is transformed into a given cluster with the unit probability, the growth is deterministic \cite{dm01c,brv01,jkk02,dgm01}. 


The other limiting case, in which an edge with the probability $1-p \to 1$, $1-p \neq 1$ is transformed into itself, we arrive at more ``standard'' models of the networks growing under the mechanism of preferential linking. 
For example, the transformation shown in Fig. \ref{f1}(b) with $p \to 0$, $p \neq 0$, generates the highly clustered scale-free network with numerous loops, 
which was introduced in Ref. \cite{dm01c}. The $\gamma$ exponent of its degree distribution ($P(k) \propto k^{-\gamma}$) is $3$. 

In another example (see Fig. \ref{f1}(c)), in the limit $p \to 0$, $p \neq 0$, we obtain the Barab\'asi-Albert model with a low clustering. 

In fact, the transformations, discussed in this communication, are very similar to those which are used in a standard real space renormalization group for disordered lattice models. The technique of the real space renormalization group is well developed for disordered lattices. So, its application to these growing random networks is a standard routine. 
\\

\noindent
$^{\ast}$      E-mail address: sdorogov@fc.up.pt \\
$^{\dagger}$   E-mail address: jfmendes@fc.up.pt \\ 
$^{\ddagger}$  E-mail address: alnis@samaln.ioffe.rssi.ru

\end{multicols}


\begin{references}
\bibitem{remark4}
Preferential attachment was introduced by using a rather specific network, a citation graph: the Barab\'asi-Albert model \protect\cite{ba99}. In such graphs, new vertices become attached to old ones, and actually only one of the ends of the each edge becomes attached preferentially. In this situation the usual form of preference, $f(k)$, is the only possibility.
\end{references}
\end{document}